# Novel highest-T$_c$ superconductivity in two-dimensional Nb$_2$C MXene

*Zaheer Ud Din Babar, M. S. Anwar,* Muhammad Mumtaz, Mudassir Iqbal, Ren-Kui Zheng, Deji Akinwande, Syed Rizwan**

Zaheer Ud Din Babar, Assoc. Prof. Dr. Mudassir Iqbal, Assoc. Prof. Dr. Syed Rizwan
Physics Characterization and Simulations Lab (PCSL), School of Natural Sciences (SNS), National University of Sciences & Technology (NUST), Islamabad 44000, Pakistan.
Emails: syedrizwanh83@gmail.com, syedrizwan@sns.nust.edu.pk

Dr. M. S. Anwar
Department of Materials Science and Metallurgy, University of Cambridge, CB3 0FS Cambridge, United Kingdom
Email: msa60@cam.ac.uk

Assoc. Prof. Muhammad Mumtaz
Materials Research Laboratory, Department of Physics, Faculty of Basic and Applied Sciences (FBAS), International Islamic University (IIU) Islamabad 44000, Pakistan.

Prof. Dr. Ren-Kui Zheng
State Key Laboratory of High Performance Ceramics and Superfine Microstructure, Shanghai Institute of Ceramics, Chinese Academy of Sciences, Shanghai 200050, China.

Prof. Dr. Deji Akinwande
Microelectronics Research Center, The University of Texas at Austin, Austin, Texas 78758, United States.

Currently, superconductivity in two-dimensional (2D) materials is a hot topic of research owing to their potential technological applications. Here, we report observation of superconductivity in a 2D Nb$_2$C MXene with transition temperature of 12.5 K, which is the highest transition temperature in MXene attained till now. We systematically optimized the chemical etching process to synthesize the Nb$_2$C MXene from its Nb$_2$AlC MAX phase. The X-ray diffraction (XRD) shows a clear (002) peak indicating the successful formation of MXene as well as a significant increase in the c-lattice parameter from 13.83Å to 22.72Å that indicates the delamination of Nb$_2$C MXene sheets as revealed by morphological study using scanning electron microscope. The Meissner effect is detected using superconducting quantum interference device (SQUID: Quantum design). Lower and upper critical fields as a function of temperature follow the Ginzburg-Landau (GL) theory indicating the superconducting nature of the $Nb_2C$ MXene. Strong-electron phonon interaction and the large density-of-states at Fermi level may cause the emergence of superconductivity at such a higher transition temperature which has theoretically been predicted for $Mo_2C$ MXene. Our work is a significant advancement in the field of research and potential applications of 2D MXene.

Superconductivity in two-dimensional (2D) materials is fascinating in finding new Physics and is the foreground for many promising technological applications. Variety of astonishing physical phenomenon emerged in 2D materials, such as magnetism and superconductivity can revolutionize the field of spintronics and superconducting devices. Synthesis of new class of 2D materials, known as MXenes ($M_{n+1}X_n$), auctioned the possibility of magnetism in new 2D materials and their possible utilization in spintronics.[1-3] MXenes have attracted a lot of attention due to wealth of unusual physical and chemical properties as they were successfully derived from 'MAX' phase by selective etching of 'A' layers using hydrofluoric (HF) acid. The 'A' layer is chemically more reactive to fluoride containing acids and can be removed efficiently.[4] Such 2D transition metal carbides (TMC's) hold fascinating properties due to the coexistence of ionic, covalent and metallic bonds.[5]

Superconductivity in 2D materials hold a range of promising future applications thus, a lot of efforts are being made to synthesize 2D superconducting MXenes. Recently, 2D α-$Mo_2C$ crystals were grown using chemical vapor deposition technique with crystal thickness in nanometer range.[6] The superconductivity was observed in 2D α-$Mo_2C$ MXene crystals with $T_c$ around 3.6 K via resistivity versus temperature measurements. The suppression in $T_c$ was observed with the increase of external applied magnetic field and the superconductivity was vanished completely at $H = 1.50$ T. Although the superconductivity phenomenon observed in α-$Mo_2C$ MXene crystals is fascinating however, the main drawback is too low value of its $T_c$ (K). Further studied was carried out on Mo$_2$C crystals in order to explore its superconducting nature.[7,8] They reported higher transition temperature $T_c$~ 8.02 K. The authors however mentioned the small coherence length of 13.5 nm which was attributed to the presence of surface disorder. The enhanced superconducting properties of α-Mo$_2$C MXene crystals was proposed to be due to strain-induced coupling and increased lattice defects which were mainly associated with the Cu/Mo bilayer substrate and was not due to the intrinsic structure of MXene itself. The experimentally reported superconducting transition temperature in Mo$_2$C MXene is significantly lower than the theoretically predicted value of 13 K [9] which demands more efforts to be done on this material. **Figure 1a** shows the summary of critical temperatures reported till date in the 3D-MAX and 2D-MXene.

In this work, we report a new two-dimensional powdered Nb$_2$C MXene superconductor with $T_c$ ~ 12.5 K which is the highest value reported till date in 2D Mxenes and is nearly equal to the predicted value of 13 K. [9] We have systematically optimized etching process to convert bulk Nb$_2$AlC-MAX into a fine two-dimensional Nb$_2$C-MXene. The structural, morphological, magnetic and superconducting measurements were carried out in order to confirm its intrinsic superconducting nature. These results indicate that our as-synthesized powdered Nb$_2$C-MXene is superconducting indeed with highest superconducting properties in MXene 2D materials.

The structural and phase purity were characterized by X-ray diffractometer (XRD, Bruker, D8 Advance, Germany) using Cu-Kα radiations. XRD spectra of powder Nb$_2$AlC-MAX and Nb$_2$C-MXene are shown in **Figure 2a**. XRD pattern of MAX phase is well-indexed according to hexagonal P6$_3$/mmc by JCPDF 00-030-0033.[10] The disappearance of 100% intensity peak at 38.8° (00l) indicates the successful removal of 'Al' and conversion of MAX into two-dimensional MXene. In our results, peak at 12.9° is broadened and shifted to lower angle of 7.7° which shows a prominent increase in c-lattice parameter of hexagonally stacked layered structure after etching. The M-C bond has mixed characteristics of metallic/ionic/covalent bonds while M-Al is metallic.[11] Due to weak nature of M-Al bonds than M-C bonds, the treatment of MAX phases in the presence of Fluoride containing acids at high temperature results in the removal of 'Al' layer. The replacement of metallic bonds among M-Al layers is altered by weaker hydrogen bond which grants the facile separation of the sheets after HF-treatment and water intercalation during washing. This suggests that difference in relative strengths of M-Al to M-C bonds allows the selective removal of 'Al' layer without disturbing M-C layers.[12] The elemental composition of the compounds were analyzed by energy dispersive X-ray (EDX) spectroscopy as shown in **Figure 2b**. Elemental mapping indicates that the 'Al' is much reduced in MXene compared to the MAX bulk part indicating our successful etching process. There are also some signatures of presence of O and F in MXene as functional groups which are unavoidable during etching process. The scanning electron microscope (SEM) images at different resolutions were recorded in field emission electron microscope (FESEM, VEGA3-TSCAN) operated at 20 KV. The SEM images of $Nb_2AlC$ and $Nb_2C$-MXene are shown in **Figure 2(c & d)**, respectively. SEM images exhibit typical morphology of layered MXene with morphology similar to the exfoliated graphite.[13] It can be seen that the MAX is composed of bulk-like closed sheets replaced by much wider sheets after chemical reduction. After etching under optimized temperature, the MXene sheets were considerably intercalated so as to be called the two-dimensional MXene sheets.

Since our $Nb_2C$ MXene samples are in powder form so in order to study superconductivity, we investigated magnetic properties instead of electronic transport properties as the powder grains cannot be continuous to give realistic transport measurement. The magnetic measurements were carried out by using SQUID Magnetometer (Quantum Design, MPMS). The magnetization vs. temperature ($M(T)$) curves with field-cooled (FC) and zero field-cooled (ZFC) of as-synthesized $Nb_2C$ MXene powder were carefully measured under applied magnetic fields of 10 mT, as shown in **Figure 3a**. Note that, before measurements, we removed residual magnetic field by resetting the superconducting magnet of SQUID and oscillating magnetic field from 5 T to zero after linear increase. Diamangetic transition in both, ZFC and FC curves is clearly observed that suggests the existence of superconductivity with onset transition temperature $T_c^{onset}$ (K) ≈12.5 K. Expectedly, the difference between the ZFC-FC curves below $T_c$ (see **Fig. 3a**) attributes the confinement of magnetic flux that results the flux pinning and the presence of substantial mixed state which asserts that the as-prepared $Nb_2C$-MXene is a type-II superconductor.[14] There is a gradual decrease in magnetization upon decrease in the temperature that can be attributed to the formation of vortices below $T_c$ (K). The broaden superconducting transition width (ΔT) in M-T curves is another indicative of vortex state which confirms type-II superconductor.

For more detailed confirmation of emergence of superconductivity in our samples, we studied Meissner effect by measuring magnetization loops ($M(H)$) at 5 K, 10 K and 20 K as shown in **Figure 3b**. Below $T_c$, the $M(H)$ loops reveal the superconducting behavior and at 20 K, strong pramagnetic behavior is clearly present. **Figure 3c** shows $M(H)$ loops measured from zero to maximum positive field to calculate the lower critical field $H_{c1}$ and upper critical field $H_{c2}$. Both, $H_{c1}$ and $H_{c2}$ are monotonically decreasing with increase in temperature. We subtracted the pramagnetic contributions by removing linear increase at higher fields by presuming $M_d(H)=M_t(H)-M_p(H)$, where $M_d$, $M_p$ and $M_m$ are diamagnetic, pramagnetic and measured magnetizations, respectively. We calculate as $M_p=GH$, where $G$ is the gradient at higher magnetic field and $H$ is the applied magnetic field.

It can be seen that the width of M-H loop is much wider at 5K compared to other temperatures which can be attributed to the diamagnetic nature of $Nb_2C$ sample and the improved grain connectivity. This width of M-H loop decreases with increase in measuring temperature which is consistent to the fact that the superconducting nature decreases at elevated temperatures. While moving from higher temperature to lower temperature after the onset of superconducting phase, the vortices come into action that induce strong superconducting diamagnetism at low temperatures. Moreover, it can also be observed that the diamagnetic nature of the material changes to paramagnetic upon increasing the temperature above critical temperature. The vortices at low temperature are stationary and magnetic flux pinning among them becomes stronger, exhibiting strong superconductivity. Further increase in temperature may destabilize the grain connectivity, consequently destroying the flux pinning among the grains at temperature above $T_c$ hence, the material comes to the normal state.[15] The enhanced vortex dynamics at elevated temperatures above $T_c$ results in destruction of magnetic flux configurations causing lesser flux pinning among the grains thus, the material turns to normal (paramagnetic) state from the superconducting state.

The variation of $H_{c1}$ and $H_{c2}$ as a function of temperature $T$ is shown in **Figure 3d**. The solid lines show the fitting of $H_{c1}$ and $H_{c2}$ according to the Ginzburg-Landau (GL) theory for superconductors as follows: $H_{c1}(T) = H_{c1}(0)[1-(T/T_c)^2]$ and $H_{c2}(T) = H_{c2}(0)[1-(T/T_c)^2]$, respectively.[16] The values of $H_{c1}$ and $H_{c2}$ were calculated from experimental data as illustrated in **Figure 3c**. The calculated values are well-fitted to theoretical predictions of well-known Ginzburg-Landau theory and phase diagram of **Figure 3d** translates the characteristic behavior of type-II superconductor with well-defined normal (line above $H_{c2}$), vortex states (between $H_{c1}$ and $H_{c2}$) and superconducting states (below fitting line of $H_{c1}$). We also estimated the superconducing coherence length ($\xi_s$) using the GL expression $H_{c2}(0)=\varphi_0/2\pi\xi_s^2$, where, $\varphi_0$ is the flux quantum and $H_{c2}(0)$ is estimated with extra polation of data presented in **Figure 3d**. It yeilds that $\xi_s$=25 nm, which is not significantly different than that of $Mo_2C$ MXene.[17]

Now, we discuss the relativly higher $T_c$ of our samples by presuming that it may come from some imurity phases (if any) that may present in our samples such as Nb, NbC, NbAlC, Al and $Nb_2C$ (3D), as well. It is well know that $T_c$ of Nb, NbC and Al is 9.2 K, 9 K and 1.2 K, respectively. Superconductivity was also observed in NbAlC with Tc of 0.44 K. Furthermore,

G. F. Hardy *et al.* prepared 3D hexagonal, non-MXene $Nb_2C$ samples (c-LP = 4.970Å) and reported the superconductivity with $T_c$ = 9.18 K.[18] Note that none of the above materials were in two-dimensional phase. Our $Nb_2C$ sample is a 2D MXene with much higher *c*-lattice parameter (c=LP ~ 22.7Å) and $T_c$ ~ 12.5 K. It suggestes that supercondutivity in our samples is mainly emerged due to the intrinsic two-dimensional structure of $Nb_2C$ MXene. Interestingly, $T_c$ of our samples is almost close to the theoretical prediction of $T_c$ for $Mo_2C$ MXene which suggests that high $T_c$ in our sample is due to the 2D structure itself and not due to any impurity phase. The absence of impurity phases dicussed above were also confirmed by XRD results and can safely be ruled out.

The first principles density functional theory (DFT) calculations of $Mo_2C$ monolayer predicte that the $T_c$ can be enhanced to 13 K depending upon th electron-phonon coupling strength.[9] Although the theory predicted $T_c$ for $Mo_2C$ MXene but we believe it could be generalized to other MXene family with n=1 in its general formula '$M_{n+1}X_n$'. These calculations show that several bands have crossed the Fermi level with substantially higher density-of-states (DOS), which can enhance $T_c$. It was also pointed out that the prerequisite for superconductivity in all considered monolayers is the presence of metallic elements. Fortunately, the superconducting transition temperature in our $Nb_2C$ MXene system is very close to the predicted value of 13 K for $Mo_2C$ system. Moreover, in our system, all elements are metallic in nature which fulfills the condition necessary for the occurence of superconductivity. Therefore, it can be understood that the electron-phonon interaction in our $Nb_2C$ MXene system is very strong and we can predict that the DOS above the Fermi level will be much higher compared to all other 2D MXenes.

Critical current density ($J_c$) is one of the important parameters of superconductors from applications point of view. The Bean's model can be used to calculate the $J_c$ using parameters taken from $M(H)$ loops measured at different temperatures 5 K and 10 K[19-21].

Since, the sample was placed in a cylindrical glass voil hence, Bean's model for cylindrical sample was used and is given by: [22]

$$J_c = 30 \frac{\Delta M}{d}$$

Where, $\Delta M = [|M+|-|M-|]$ measured in emu/cm$^3$ and '$d$' is the diameter of the cylinder in cm. As the dimension of sample is same so, $J_c$ is directly proportional to $\Delta M$ (i.e. width of the $M(H)$ loop). The $J_c$ versus applied magnetic field ($H$) at two different temperatures (5 K and 10 K) is drawn in **Figure 4.** It can be seen from these graphs that $J_c$ was decreased with increasing temperature and also with increasing applied magnetic field.[23]

**Conclusion**

In summary, we have successfully synthesized high quality Nb$_2$C 2D MXene sheets under optimized conditions. The XRD pattern shows a significant increase in the lattice parameters indicating large intercalation of MXene sheets. The $M(T)$ and $M(H)$ measurements indicate the existence of superconductivity in as-synthesized MXene powder with high $T_c$ =12.5 K. The fitting of critical fields ($H_{c1}$ and $H_{c2}$) as a function of temperature according to GL theory, also confirms the characteristic behavior of type-II superconductivity. It is worth paying attention that, to our knowledge, this is the first study on the superconductivity of as-prepared MXene (Nb$_2$C-powder in present case) which shows highest $T_c$ =12.5 K, which is clos to the theoretically predicted value of 13 K. The possible reasons for such a higher $T_c$ in Nb$_2$C MXene maybe strong-electron phonon interaction as well as the presence of higher density-of-states at the Fermi level. This work is a significant development to explore the physical properties exhibited by 2D materials particularly the MXenes which offer unique potential for future applications.

**Experimental Section**

The materials used for etching process includes Hydrofluoric acid solution (50 wt. % in H$_2$O, ≥99.99%), Niobium Aluminum Carbide Powder (Nb$_2$AlC, purity 90-95 %, 200mesh), de-ionized water, absolute ethanol. In order to obtain the two-dimensional sheets of MXene, etching was performed to selectively remove aluminuim (Al) from $Nb_2AlC$-MAX Phase. The $Nb_2AlC$ powder was added in 50% HF (sigma Aldrich) in the ratio (1:10) at room-temperature (RT) in a Teflon beaker and was continuously stirred for 90 hours by Teflon-coated magnetic stirrer.[24] Etching of our MAX was not successfully done at such conditions. Then, we performed a systematic study of etching process at optimized time of 40 hrs and at varying temperature (45°C - 85°C) to get our final product. After HF treatment, the suspension was washed several times by deionized water and centrifuged at 3500 rpm for 5 minutes until the pH reached to 7. The HF selectively removes the Al, which, in succession replaced by F, O and/or OH. Each time, the supernatant was removed and settled powder was removed from centrifuge bottles using ethanol and dried at room-temperature.

The schematic of this etching process is shown in **Figure 1b** which gives a physical insight towards sheets separation upon HF treatment and separation of finally obtained sheets after washing and drying. The yield here is addressed as the weight of MXene/ MAX times 100 ($W_{MXene}/W_{MAX} \times 100$), was, almost 100% and resultant MXene was obtained in maximum amount without prominent losses. For magnetic measurement the powdered sample was put into a capsule mounted in a tube. Before starting the measurement, the system was calibrated by using an oscillatory field in order to eliminate the entire stray fields among the coils, so the real magnetic behavior of the sample could be observed.

**Acknowledgements**

The authors are thankful to Higher Education Commission (HEC) of Pakistan for providing research funding under the Project No.: 6040/Federal/NRPU/R&D/HEC/2016 and HEC/USAID for financial support under the Project No.: HEC/R&D/PAKUS/2017/783.

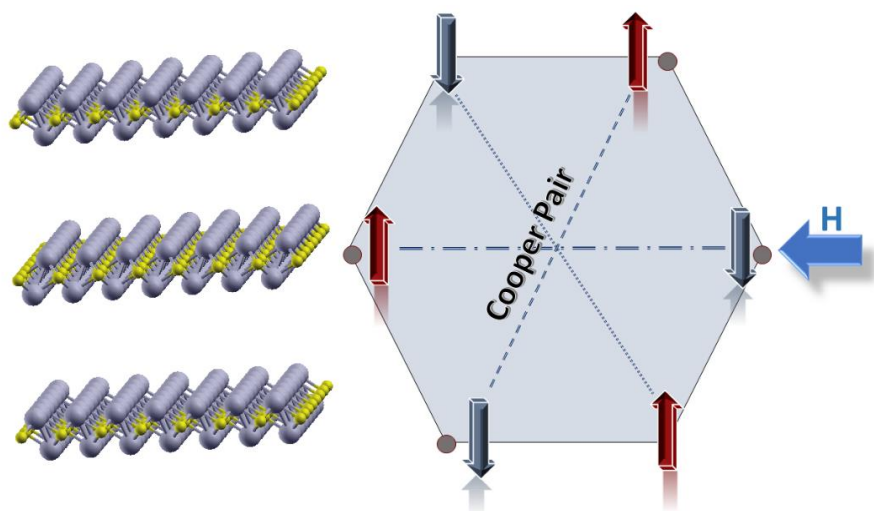

**Table of Content (TOC) – Art File**

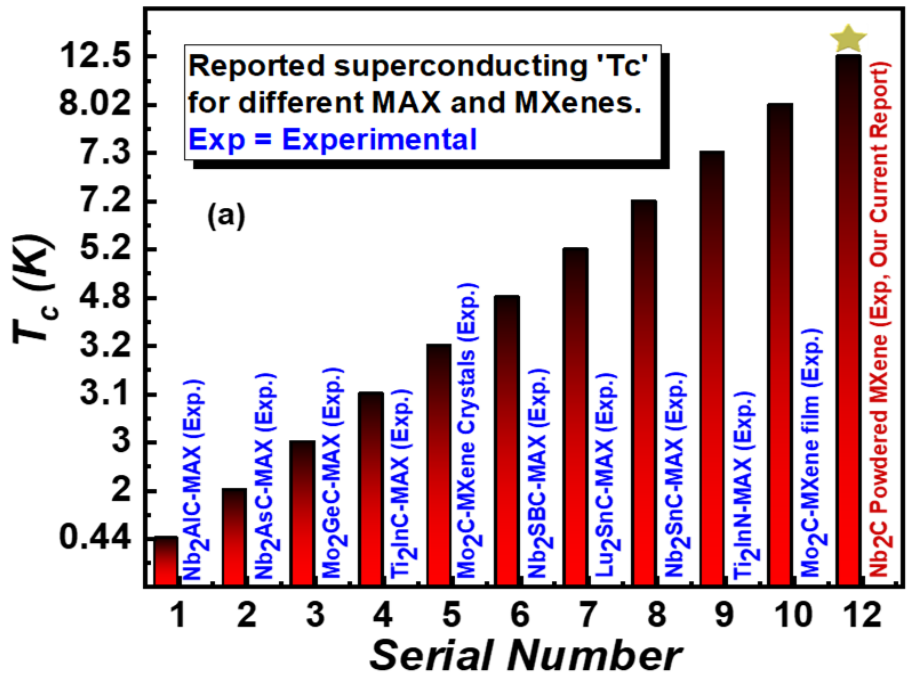

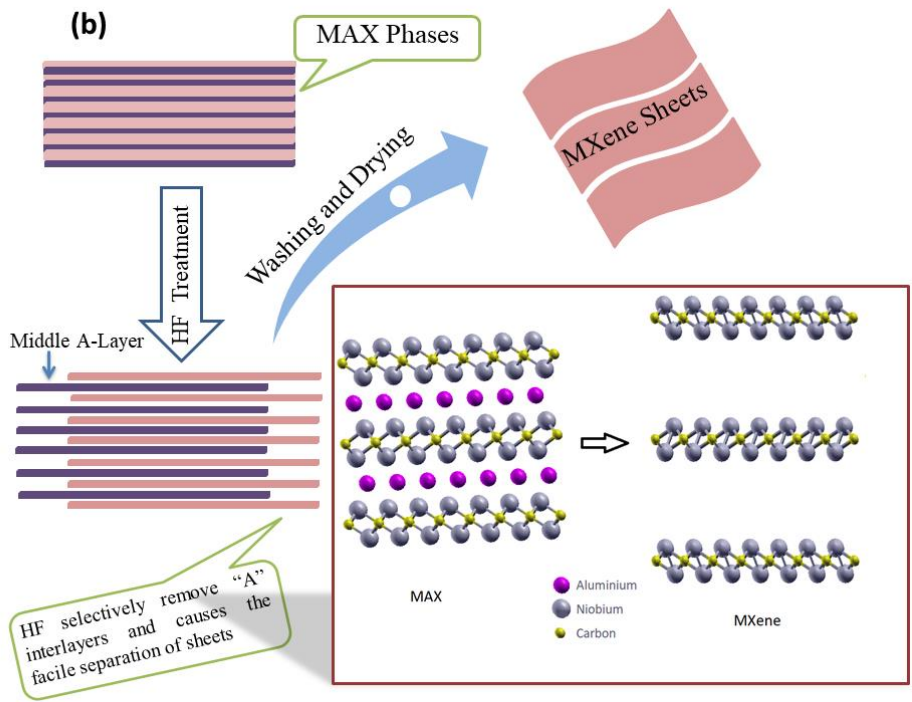

**Figure 1.** (a) Summary of superconducting transition temperature of various MAX (3D) and MXene (2D) superconductors reported till date. (b) Schematic of chemical etching protocol of synthesis of MXene 2D sheets from MAX phase.

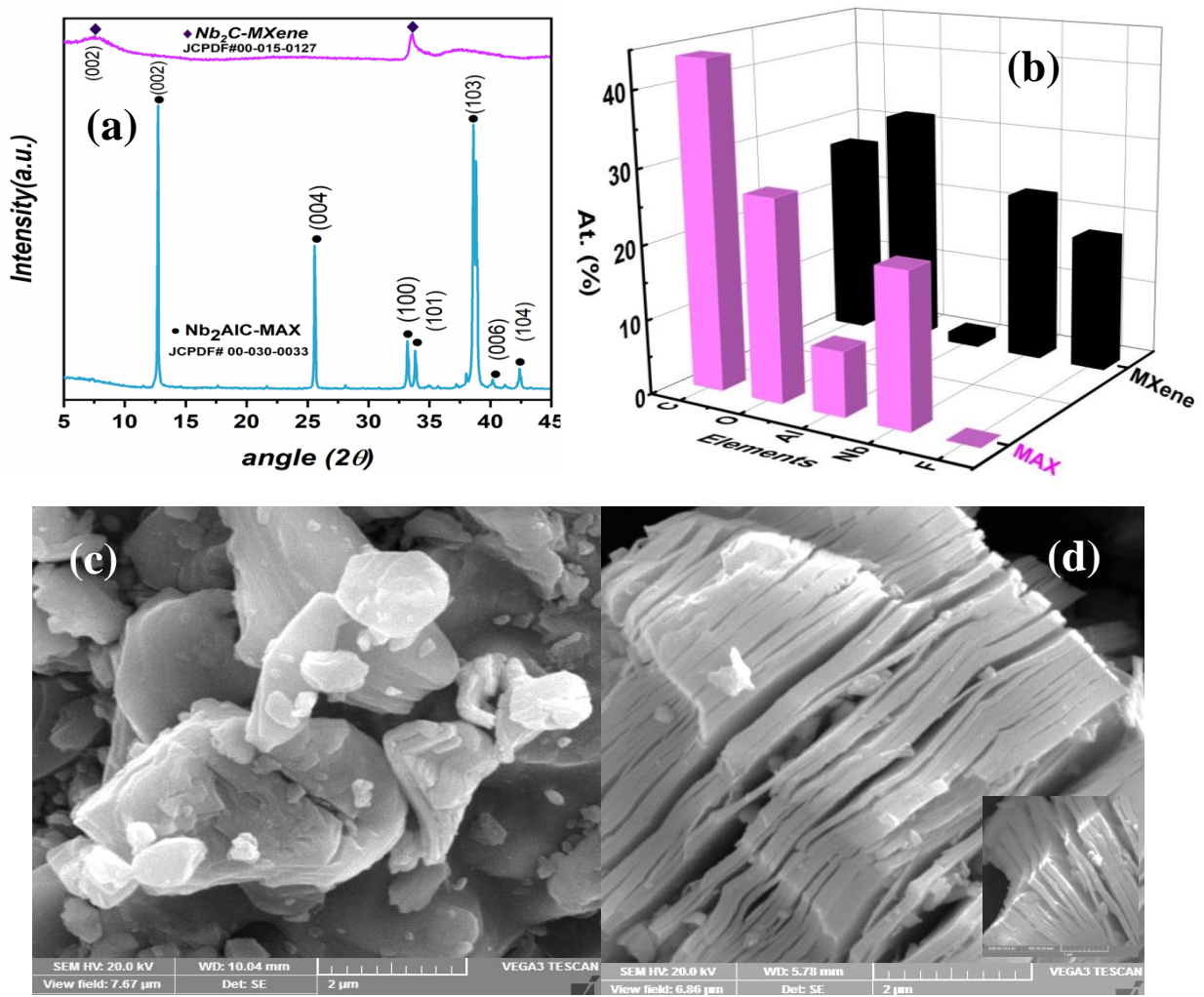

**Figure 2.** (a) X-ray diffraction of Nb$_2$AlC MAX and MXene before and after treatment, respectively, indicating the removal of 'Al' (b) energy dispersive X-ray (EDX) spectrum of MAX and MXene which shows significant decrease in 'Al' percentage after HF treatment (c) Scanning Electron Microscopy (SEM) shows lamellar structure of MXene.

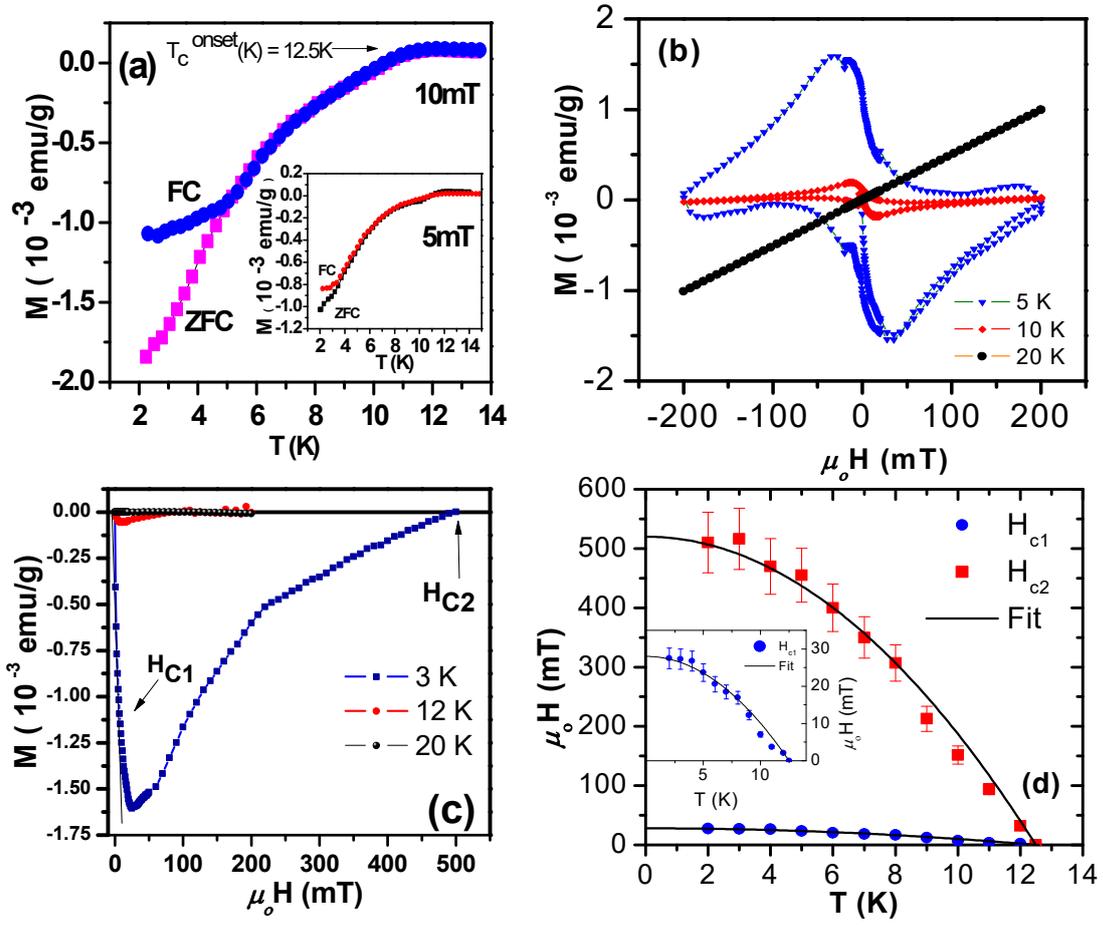

**Figure 3.** (a) zero-field-cooled (ZFC) and Field-cooled (FC) curves indicating $T_c^{onset}$ (K) =12.5 K (b) M-H loops (M vs. H) shows typical loops Type-II superconductor (c) Lower critical field ($H_{c1}$) and Upper critical field ($H_{c2}$) depicts the characteristic fields of Type-II superconductor (d) fitting of experimentally calculated $H_{c1}$ and $H_{c2}$ as a function of temperature $T/T_c$ in accordance with GL theory of superconductivity.

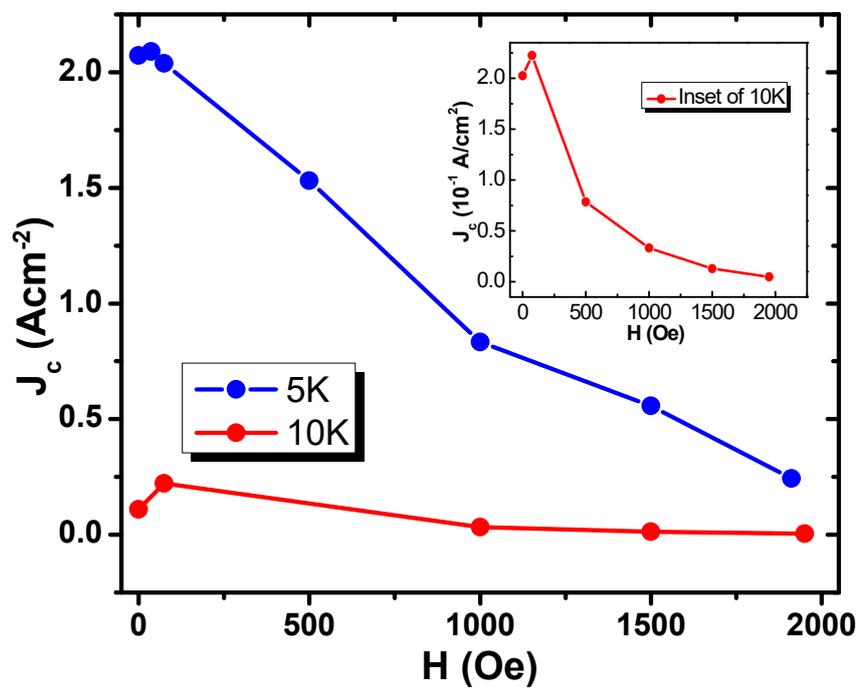

**Figure 4.** Calculated critical current density $J_c$ versus applied magnetic field at 5 K and 10 K: inset is the zoom-version of the same at 10 K.